# Ventilation regime in a karstic system (Milandre Cave, Switzerland)

Julia GARAGNON[1,2,3], Marc LUETSCHER[1] & Eric WEBER[1]

(1) Swiss Institute for Speleology and Karstology (SISKA), Rue de la Serre 68, 2300 La Chaux-de-Fonds, Switzerland
(2) Laboratoire des Sciences du Climat et de l'Environnement, UMR8212, CEA-CNRS-UVSQ, Bat. 714, Ormes des Merisiers, 91191 Gif-Sur-Yvette, France.
(3) Laboratoire Environnements et DYnamiques des Territoires de Montagne, UMR5204, CNRS-USMB-UGA, Bat. Pôle Montagne, Campus Scientifique, 73376 Le Bourget du Lac Cedex, France.
julia.garagnon@gmail.com (corresponding author)

## Abstract

Cave climatology and its impact on contemporary biogeochemical cycles are still poorly documented. Ventilation in karst environment plays a fundamental role in these two fields and its understanding could bring elements to study them. However, only a few cavers have tried to understand and describe it, very often in a qualitative way or by theoretical approaches. The aim of this study is to test physical concepts with empirical data. For this purpose, a ventilation model has been built and compared with field temperature and air velocity measurements in the Milandre Cave Laboratory (Switzerland). The model explains about 95% of the measured airflow thus confirming the major role of temperature on the air dynamics. However, these first results also reveal that the measured winter air flow is lower than predicted by the model and that the air flow reversal occurs at a lower temperature than anticipated. Combined with a forced ventilation experiment these results underline the influence of the atmospheric composition (particularly the water vapor and concentration in $CO_2$ and $O_2$), water flow rates and network geometry on the air flow. This work paves the way for a better quantification of heat and mass fluxes in relation to underground ventilation.

## Résumé

**Régime de ventilation en milieu karstique (grotte de Milandre, Jura suisse)**. La climatologie et l'étude des cycles biogéochimiques contemporains sont peu documentés à ce jour. La ventilation en milieu karstique joue un rôle fondamental dans ces deux domaines et sa compréhension pourrait apporter des éléments pour les étudier. Toutefois, seuls quelques spéléologues se sont attelés à la comprendre et la décrire, bien souvent de manière qualitative ou par des approches théoriques. Le but de cette étude est de tester ces théories en se basant sur des données mesurées. Pour cela, un modèle de ventilation a été construit et confronté aux mesures de terrains (température et vitesse de l'air) effectuées sur le site de Milandre (Suisse). Le modèle explique environ 95% des débits mesurés confirmant le rôle prépondérant de la température sur le courant d'air. Toutefois, ces résultats mettent également en évidence des débits d'air hivernaux mesurés plus faibles que ceux prédits par le modèle et une inversion du flux d'air pour une température extérieure plus faible que celle attendue. Combinés à une expérience de ventilation forcée, ces résultats montrent l'influence de la composition atmosphérique (en particulier la concentration en $CO_2$, $O_2$ et vapeur d'eau), des débits d'eau et de la géométrie du réseau sur le courant d'air. Ce travail ouvre des perspectives pour une meilleure quantification des flux de chaleur et de masse en milieu karstique en lien avec la ventilation souterraine

## 1. Introduction

The air flow in karstic networks has always been considered by cavers as a guide for speleological explorations but has gained less attention in scientific studies. Yet, ventilation influences numerous parameters including the cave temperature, the gas composition of the air and the mineralization of water (LISMONDE, 2002). A change in these parameters has impacts on the speleothem fabrics, water pH, biotopes and cavers health (e.g. MATTEY *et al.*, 2009; SPOTL *et al.*, 2005; MAMMOLA *et al.*, 2019). Hence, studying ventilation is key for a better understanding of heat and mass transports in caves and their impact on subsurface ecosystems and paleoenvironmental records. Only few cavers and scientists have described air circulations in karst systems either qualitatively or with physical concepts (e.g. BADINO, 2010; LISMONDE, 2002). With the recent developments in instrumentation, these initial models can now be tested and improved based on robust empirical data.

Many caves present two or more entrances and function as wind tubes, i.e., a network with an upper and a lower entrance which induce a forced air convection (LISMONDE, 2002). The process results from a difference of the pressure which reflects the difference of density between two air columns, one inside the cave and one outside. This is directly related to the ideal gas law (LISMONDE, 2002).

This article presents the firsts results from the study of air flows in the karstic system of Milandre

## 2. Materials and methods

Located in NW Switzerland, along the Swiss-French border in the Swiss tabular Jura (Ajoie), the Milandre cave network is a natural cave laboratory (Fig.1) which has been extensively monitored over the last 20 years (KARST, 2017). This 10.5km network presents two artificial entrances, a historical touristic entrance in the downstream part (402m.asl), and a 40m deep shaft system in the upstream part of the cave (509m.asl) (GIGON & WENGER, 1986) (Fig.1). The average cave temperature of 10.5°C contrasts with external temperatures inducing a strong chimney effect with ascending air flow in winter and descending in summer.

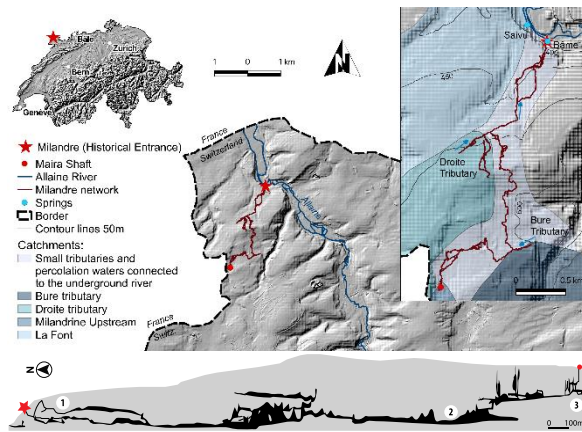

*Figure 1: Site location and projected profile (vertical exageration x2) of the Milandre Cave Laboratory; 1)Mil_Fortin, 2)Mil_Shunt, 3)Mil_Amont.*

The *Milandrine* cave river flows over c. 4km along the main conduit. It is fed by a 13.3km² catchment area (JEANNIN (1996)) encompassing two main tributaries: Bure and Droite (Fig.1). The system is drained toward three main outlets, among which, the perennial resurgence of the Saivu (375m. asl, [20-200]l/s) and the overflow spring of Bame [200-2000]l/s (VUILLEUMIER, 2018) (Fig.1).

The cave monitoring includes, among others, four hydrological stations and three wind stations located in the upstream- (Mil_Amont), downstream- (Mil_Fortin) and middle- (Mil_Shunt) part of the cave. Two different types of instruments 1) hot-wire anemometer and 2) flowmeters provide an estimate of the airflow after integration with the conduit section. The average cave temperature, measured in the stream ("oiler" probe) between 2015 and 2018, is 10.5°C ±0,1°C. In addition, data from a nearby MeteoSwiss station (Fahy) have been collected.

The aeraulic model assumes that ventilation is dominated by the temperature difference between the inside and outside atmosphere following the relation $Q_{air} = a\sqrt{\Delta T}$. The data have been filtered as $Q_{air}^2 = ]0.01; 11.391[$ to eliminate instrumental outliers and artifacts, centered at 0 and positive $\Delta T$ have been treated separately from negative $\Delta T$ to consider the difference between winter and summer regime.

To determine the influence of the network geometry on cave ventilation the upstream and downstream cave entrances were successively closed. This experiment took place over a stable and warm meteorological period ($T_{mean}$ 22°C) from July 20, 2020 to August 15, 2020 (Fig.2).

| Date | Time (day) | Maira | Fortin | Managers |
|---|---|---|---|---|
| Until 07/30 | - | Open | Open | JG-CP |
| 07/30/20 | 5 | Close | Close | JG-CP |
| 08/05/20 | 3 | Close | Open | JG-CP |
| 08/09/20 | 5 | Open | Close | JG-CP |
| 08/15/20 | - | Open | Open | SCJ |

*Figure 2: Opening-closing doors experimentation program.*

Parallel to the airflow monitoring, temperature, $CO_2$ and $O_2$ concentration were measured at fixed intervals (15-30min) as well as manually. Because of an instrument failure on July 10, 2020 at the downstream station (Mil_Fortin), only sporadic airflow measurements are available.

## 3. Results

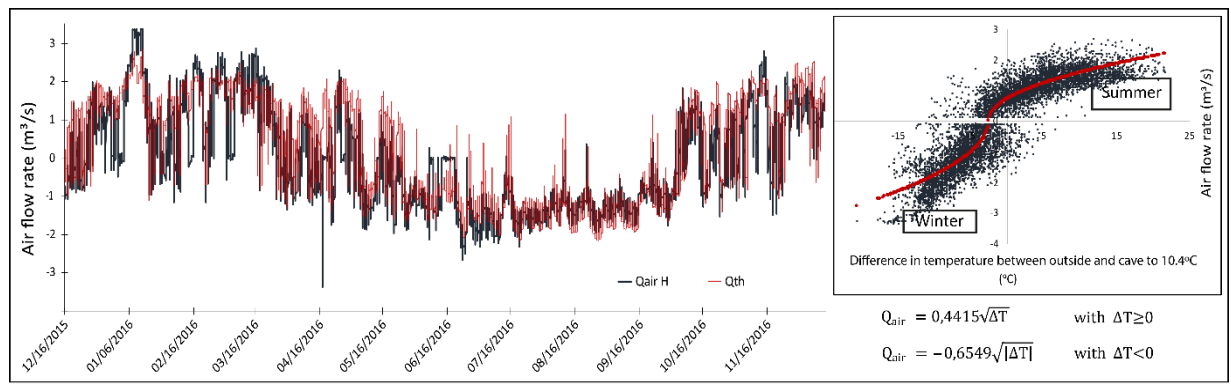

*Figure 3: Continuous airflow recorded (grey) at Mil_Fortin from Dec.2015 to Dec.2016 and modelled values (red).*

Airflow measured over 1 year (from December 2015 to December 2016; Fig. 3) in the downstream part of the cave (Mil_Fortin) shows daily oscillations and a seasonal signal marked by a flow reversal in the fall and spring. The summer months seem to be more stable than the winter ones, where maxima airflows (3m³/s) are observed. Under turbulent regimes, airflow is proportional to the square root of the temperature difference between the cave and the outside air (LISMONDE, 2002; LUETSCHER, 2005) and our modelled values show an excellent correlation with the airflow measured at Mil_Fortin ($R^2$ = 0.9, Fig.3).

The residual airflow curve (Qair measured – Qair modelled) (Fig.4) suggests that the model explains 95.4% of the data, supporting that temperature is the dominant parameter controlling ventilation. The remaining 4.6% correspond mostly to measured flows lower than predicted. This suggests the influence of one or more parameters, other than temperature, on the air flow.

Figure 4 shows that the decrease in the airflow (in red) correlates with flood events during the winter season. The correlation is, however, less obvious during the fall.

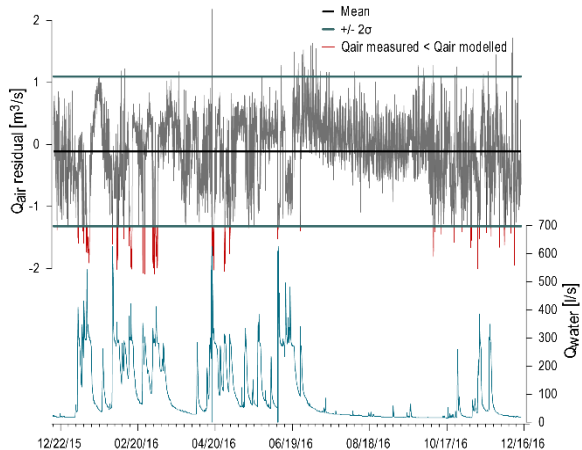

*Figure 4: Residual airflow curve (grey/red) and water flow (blue).*

The three major floods visible during that time-period do not explain the high variability of the air flow.

The results of the forced ventilation experiment (Fig.5) show that the air flow measured at Mil_Shunt does not react to the opening and closing of the doors and, rather, follows the modelled values. In contrast, the experiment clearly reveals the impact of the ventilation on the $CO_2$-$O_2$ concentration in the cave, both measured downstream and upstream.

The closing of the doors on July 30, 2020 at 5:35 p.m. at Mil_Fortin and 8:38 p.m. at Mil_Amont, interrupted the airstream and led to an increase in $CO_2$ levels at Mil_Fortin simultaneous to a drop in $O_2$. Upstream, the $CO_2$ concentration stabilized until the reopening of the Mil_Amont door. The Mil_Fortin opening on August 5, 2020 at 6:21 p.m. (upstream kept closed) resulted in a recovery of the air current to 1.6m³/s downstream and a sharp drop in $CO_2$ concentration due to the ventilation of the system. The closure of Mil_Fortin door on August 9, 2020 at 14:26 and the reopening of the Maira the same day at 19:42 didn't seem to have any influence on the downstream concentration in $CO_2$ which continued to decrease (and conversely the $O_2$ to increase) until it stabilized. However, an intensification of the airflow was observed in the "muddy passage" downstream Mil_Fortin, increasing from 1.2m³/s to 1.8m³/s. An anti-correlation between $O_2$ and $CO_2$ at Mil_Fortin and a $CO_2$ peak caused by the temporary opening of Mil_Amont on August 5th a $CO_2$ is also noted.

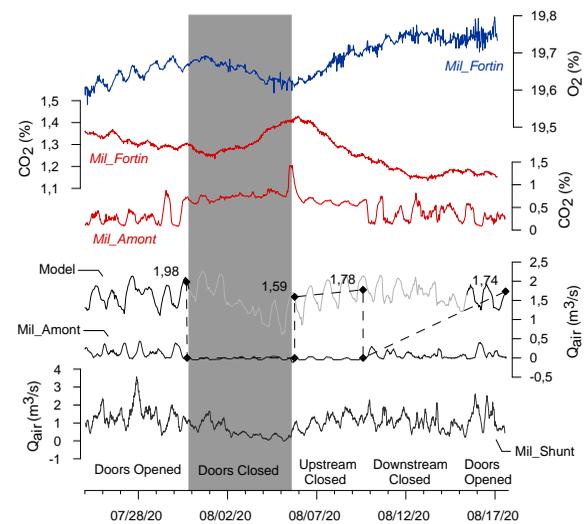

*Figure 5: Modelled airflow in normal conditions (black & grey curve) and spot airflow measurements at Fortin (dashed curve) during the experiment. Continuous airflow records at Mil_Amont and Mil_Shunt (black). $CO_2$ values (red) at Mil_Fortin and Mil_Amont, and $O_2$ (blue) at Mil_Fortin.*

## 4. Discussion

The scattered data in Figure 3 reveal a hysteresis phenomenon which is interpreted as being related to the thermal inertia of the cave walls (LISMONDE, 2002). In our quasi-static model, the temperature of the cave is set as constant, and does not take into account the seasonal temperature changes of the rock in the entrance areas. Considering an average at 10.5°C is thus a simplification of the model.

The cloud shift of -2.5°C (Fig.3) suggests that the air flow reversal does not occur when the outside air temperature equals the temperature of the cave, i.e 10.5°C, but rather when the outside temperature is 8°C. This means that the temperature is not the only parameter controlling the air flow reversal. The difference in weight between the external and internal air masses, is at the origin of the air stream. The density of these air columns is dominated by temperature but also depends on the gas composition of the air, especially water vapor, $CO_2$ and $O_2$ (LISMONDE, 2002). In particular, the relative humidity in a cave is generally close to 100% and the concentration of $CO_2$ in the air at Milandre is typically between 1 and 3%, versus 0.04% in the external atmosphere. According to Fig. 6, for an atmospheric pressure of $10^5$ Pa, the density of a column of indoor air at 10.5°C, 100% humidity, 1.5% $CO_2$ and 19.5% $O_2$ is 1.245kg/m³. A similar density corresponds to an outside air column at 8°C, 70% humidity and 0.04% $CO_2$. The influence of humidity, $CO_2$ and $O_2$ is therefore significant at Milandre and explains the 2.5°C offset observed in the model.

The importance of humidity is further explained by the geometry of Milandre. The upstream part of the cave presents a positive thermal anomaly associated with a relative humidity close to saturation all year long. This anomaly has all the more influence on the density of the air column as the upstream part presents a sharp drop in elevation.

| Temperature [ºC] | Humidity [%] | CO2 [%] | O2 [%] | Density [kg/m3] |
|---|---|---|---|---|
| 10.5 | 70% | 0.04% | 21% | 1.235 |
| 10.5 | 100% | 1.50% | 20% | 1.245 |
| 10.5 | 100% | 2.00% | 19% | 1.248 |
| 8 | 100% | 2.00% | 19% | 1.257 |
| 8 | 70% | 0.04% | 21% | 1.245 |
| 8 | 50% | 0.04% | 21% | 1.244 |

*Figure 6: Density variation according to different parameters*

The decrease of the airflow in winter in case of flooding is attributed to the opposite direction between the ascending winter air and the dragging of air by the descending water flow (Fig.7, n°2) intensified by overpressure wave effects (Fig.7, n°1), and water piston effect (Fig.7, n°3) (MANGIN & ANDRIEUX, 1988).

The arrival of cold water could also imply a brief decrease in the thermal gradient and thus a weakening of the air flow. However, the first comparisons between $T_{water}$ and $T_{air}$ in Mil_Amont, Mil_Shunt and Saivu between September 2019 to June 2020 tend to invalidate this hypothesis but would deserve a more detailed study to understand the influence of water on the cave air temperature and thus on the airflow. The high variability of the air stream in autumn shows that at least one other factor influences ventilation. A dysfunction of the device or an anthropic interaction (e.g. modification of the position of the anemometer by cavers) cannot be excluded.

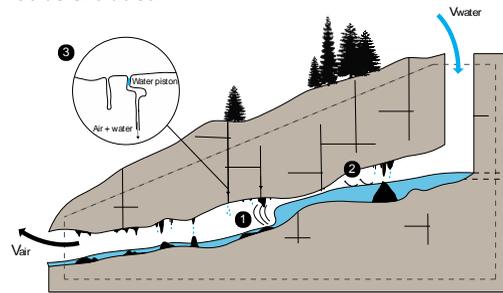

*Figure 7: Schematic illustration of the different processes that impact the airflow during a flood.*

The forced ventilation experiment underlines the influence of the geometry of the cave. At low-water levels (water flow <20L/s), some passages can be dewatered. This modifies the flow of the air stream which is no longer restricted to the passage with the Mil_Fortin anemometer. This one therefore measures only part of the airstream during the summer period. As the air flow at Mil_Shunt was not influenced by the experiment, it assumes the existence of an entrance at a higher elevation than the Maira shaft.

## 5. Conclusion

As the airflow modelling in Milandre explains 95% of the measured airflow, the major role of temperature on the air dynamics is confirmed. The model suggests a non-negligeable influence of humidity and $CO_2$, resulting in a -2.5°C offset on the airflow reversal temperature, initially expected at 10.5°C (average cave temperature). Water levels also have a marked impact on the air flow, with a decrease in winter, during flood events, and a change in the flow distribution at low-water level. Finally, this study also highlights the role of ventilation on the transport of $CO_2$-$O_2$ calling for a better understanding of carbon fluxes in karstic environments.

## Acknowledgments

*We gratefully thank Claudio PASTORE and Félix ZIEGLER for their help with the monitoring. We would also acknowledge the long-term contribution of the Group Karst of A1*

## References

BADINO G. (2010). Underground meteorology - What's the weather underground? *Acta Carsologica*, *39/3*, 427-448.

GIGON R. & WENGER R. (1986). *Inventaire Spéléologique de la Suisse - Tome 2 - Canton du Jura*. Commission de Spéléologie de la Société suisse des Sciences naturelles, La Chaux de Fonds, 292.

JEANNIN P.Y (1996). *Structure et comportement hydraulique des aquifères karstiques* [PhD Thesis]. CHYN, 237.

KARST G. (2017). *A16-Section 2: Etude d'impact sur la grotte de Milandre. Rapport final*. Canton du Jura, service des ponts et chaussées.

LISMONDE B. (2002). *Climatologie du monde souterrain. Tome 1 et 2*. Comité départemental de Spéléologie de l'Isère, 168, 362.

LUETSCHER M. (2005). *Processes in ice caves and their significance for paleoenvironmental reconstructions* [PhD Thesis]. University of Zurich, 154.

MAMMOLA S., PIANO E. & CARDOSO, P. (2019). The effects of global climatic alterations on cave ecosystems. *Anthr. Rev.*, 6, 98-116.

MANGIN A. & ANDRIEUX C. (1988). Infiltration et environnement souterrain, le rôle de l'eau sur les paramètres climatiques. *Actes des Journées Félix Trombes*, *T 1*, 78-95.

MATTEY D. P., FAIRCHILD I. J. & ATKINSON, T. C. (2009). Seasonal microclimate control on calcite fabrics, stable isotopes and trace elements in modern speleothem from St. Michaels Cave, Gibraltar. *Geochim. Cosmochim. Ac.*, *73*, A849-A849.

SPOTL C., FAIRCHILD I.J. & TOOTH A.F (2005). Cave air control on dripwater geochemistry, Obir Caves (Austria): Implications for speleothem deposition in dynamically ventilated caves. *Geochim. Cosmochim. Ac.*, 69, 2451-2468.

VUILLEUMIER C. (2018). *Hydraulics and sedimentary processes in the karst aquifer of Milandre (Jura Mountains, Switzerland)* [PhD Thesis]. CHYN, 130.